\documentclass[preprint2]{aastex}

\newcommand{\feh}{\mbox{[Fe/H]}}
\newcommand{\teff}{\mbox{$T_{\rm eff}$}}
\newcommand{\logg}{\mbox{$\log g$}}
\newcommand{\vsini}{\mbox{$v \sin I$}}
\newcommand{\mictrb}{\mbox{$\xi_{\rm t}$}}
\newcommand{\mactrb}{\mbox{$v_{\rm mac}$}}

\newcommand{\kms}{\mbox{km\,s$^{-1}$}}
\newcommand{\ms}{\mbox{m\,s$^{-1}$}}

\newcommand{\mplanet}{\mbox{$M_{\rm P}$}}
\newcommand{\rplanet}{\mbox{$R_{\rm P}$}}
\newcommand{\densplanet}{\mbox{$\rho_{\rm P}$}}
\newcommand{\mjup}{\mbox{$M_{\rm Jup}$}}
\newcommand{\rjup}{\mbox{$R_{\rm Jup}$}}
\newcommand{\densjup}{\mbox{$\rho_{\rm Jup}$}}
\newcommand{\mstar}{\mbox{$M_*$}}
\newcommand{\rstar}{\mbox{$R_*$}}
\newcommand{\densstar}{\mbox{$\rho_*$}}
\newcommand{\msol}{\mbox{$M_\odot$}}
\newcommand{\rsol}{\mbox{$R_\odot$}}
\newcommand{\denssol}{\mbox{$\rho_\odot$}}

\def\secos{$\sqrt{e} \cos \omega$}
\def\sesin{$\sqrt{e} \sin \omega$}

\shorttitle{WASP-40b: Independent Discovery of HAT-P-27b}
\shortauthors{Anderson et al.}

\begin{document}

\title{WASP-40b: Independent Discovery of the 0.6 \mjup\ Transiting Exoplanet 
HAT-P-27b}

\author{D.R. Anderson\altaffilmark{1},
S.C.C. Barros\altaffilmark{2},
I. Boisse\altaffilmark{3,4},
F. Bouchy\altaffilmark{3,5},
A. Collier Cameron\altaffilmark{6},
F. Faedi\altaffilmark{2},
G. Hebrard\altaffilmark{3,5},
C. Hellier\altaffilmark{1},
M. Lendl\altaffilmark{7},
C. Moutou\altaffilmark{8},
D. Pollacco\altaffilmark{2},
A. Santerne\altaffilmark{8},
B. Smalley\altaffilmark{1},
A.M.S. Smith\altaffilmark{1},
I. Todd\altaffilmark{2},
A.H.M.J. Triaud\altaffilmark{7},
R.G. West\altaffilmark{9},
P.J. Wheatley\altaffilmark{10}, 
 J. Bento\altaffilmark{10},
 B. Enoch\altaffilmark{6},
 M. Gillon\altaffilmark{11},
 P.F.L. Maxted\altaffilmark{1},
 J. McCormac\altaffilmark{2},
 D. Queloz\altaffilmark{7},
 E.K. Simpson\altaffilmark{2},
 I. Skillen\altaffilmark{12}
}

\altaffiltext{1}{Astrophysics Group, Keele University, Staffordshire, ST5 5BG, UK; dra@astro.keele.ac.uk}
\altaffiltext{2}{Astrophysics Research Centre, School of Mathematics \& Physics, Queen's University, University Road, Belfast, BT7 1NN, UK}
\altaffiltext{3}{Institut d'Astrophysique de Paris, UMR7095 CNRS, Universite Pierre \& Marie Curie, 75014 Paris, France}
\altaffiltext{4}{Centro de Astrof\'isica, Universidade do Porto, Rua das Estrelas, 4150-762 Porto, Portugal}
\altaffiltext{5}{Observatoire de Haute-Provence, CNRS/OAMP, 04870 Saint-Michel l'Observatoire, France}
\altaffiltext{6}{SUPA, School of Physics and Astronomy, University of St.\ Andrews, North Haugh,  Fife, KY16 9SS, UK}
\altaffiltext{7}{Observatoire Astronomique de l'Universit\'e de Gen\`eve 51 ch. des Maillettes, 1290 Sauverny, Switzerland}
\altaffiltext{8}{Laboratoire d'Astrophysique de Marseille, 38 rue Fréderic Joliot-Curie, 13388 Marseille Cedex 13, France}
\altaffiltext{9}{Department of Physics and Astronomy, University of Leicester, Leicester, LE1 7RH, UK}
\altaffiltext{10}{Department of Physics, University of Warwick, Coventry CV4 7AL, UK}
\altaffiltext{11}{Institut d'Astrophysique et de G\'eophysique, Universit\'e de Li\`ege, All\'ee du 6 Ao\^ut, 17, Bat. B5C, Li\`ege 1, Belgium}
\altaffiltext{12}{Isaac Newton Group of Telescopes, Apartado de Correos 321, E-38700 Santa Cruz de la Palma, Tenerife, Spain}

\begin{abstract}
From WASP photometry and SOPHIE radial velocities
we report the discovery of WASP-40b (HAT-P-27b),
a 0.6 \mjup\ planet that transits its 12$^{\rm th}$ magnitude 
host star every 3.04 days.  The host star is of late G-type 
or early K-type and likely has a metallicity greater than solar 
(\feh\ = $0.14 \pm 0.11$). The planet's mass and radius
are typical of the known hot Jupiters, thus adding
another system to the apparent pileup of transiting
planets with periods near 3--4 days. Our parameters match those of the recent
HATnet announcement of the same planet, thus giving
confidence in the techniques used. 
We report a possible indication of stellar activity in the host star. 
\end{abstract}


\section{Introduction}
While the Kepler mission is currently producing the most candidates for transiting 
extrasolar planets \citep[e.g.,][]{2010Sci...327..977B}, the ground-based 
transit-search programs continue to find more planets around stars at brighter
magnitudes than those found in the space missions. Of these, 
Hungarian Automated Telescope Network \citep[HATnet;][]{2004PASP..116..266B} 
and Wide Angle Search for Planets 
\citep[WASP;][]{2006PASP..118.1407P} have been the most successful.  Both 
projects are based on arrays of 200 mm f/1.8 lenses backed by CCDs, with the 
biggest difference being that HATnet operates at several longitudes while WASP 
consists of one station in each hemisphere.  
The two projects look at overlapping regions of sky, which has led to some 
near-simultaneous discoveries, such as the planet WASP-11b 
\citep{2009A&A...502..395W} also being HAT-P-10b \citep{2009ApJ...696.1950B}. 
Reporting of such independent discoveries gives important information on the 
reliability of the respective techniques and on the completeness of the 
transit surveys.

Recently HATnet announced the planet HAT-P-27b \citep{2011arXiv1101.3511B}, 
a hot Jupiter in a 3 day orbit around a $m_V$ = 12.2 star. This planet 
had been independently discovered by the WASP project and assigned the name 
WASP-40b \citep{2011EPJWC..1101004H}. 
We report here on the discovery of WASP-40b made using data from 
SuperWASP-North and WASP-South combined, together with 
radial velocities from the SOPHIE spectrograph at the Observatoire de 
Haute-Provence (OHP) observatory.

\section{Observations}
\label{sec:obs}

\begin{figure}
\centering                     
\includegraphics[height=8.4cm,angle=270]{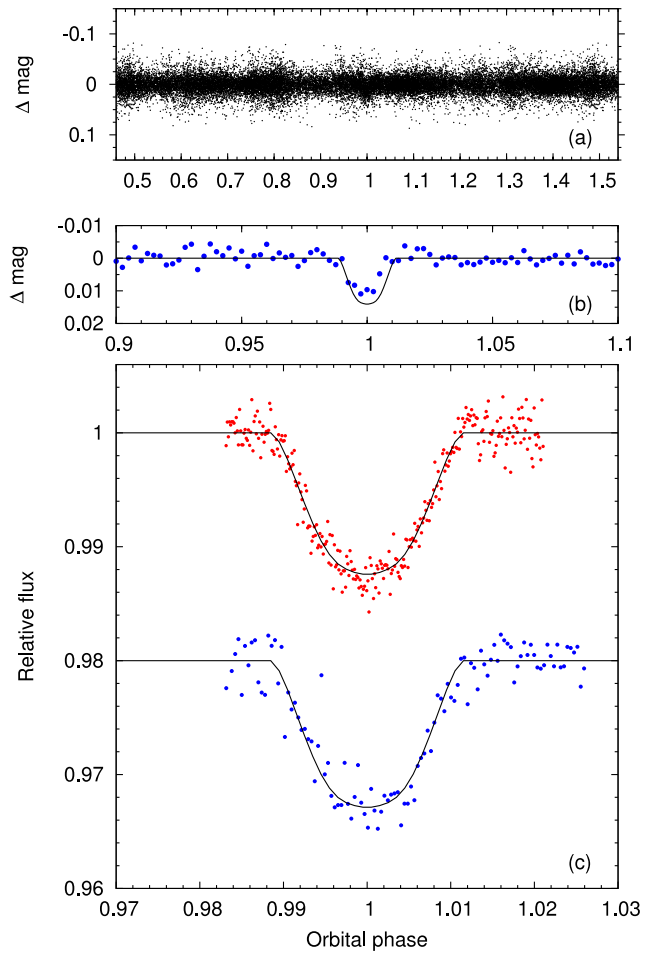}
\caption{Photometry of WASP-40, with the best-fitting transit model 
superimposed. 
{\bfseries (a)} WASP discovery light curve, folded on the ephemeris 
of Table~\ref{tab:mcmc}. 
{\bfseries (b)} WASP data around the transit, binned in time with a bin 
width of $\sim$11 minutes).
{\bfseries (c)} High-precision transit light curves from Liverpool Telescope 
RISE (top, red) and Euler/C2 (bottom, blue). 
\label{fig:phot}}
\end{figure}

We observed \objectname{WASP-40}, an $\sim$K0-type star located in Virgo,  
with the SuperWASP-North and WASP-South cameras during the three seasons of 
2008--2010. 
A transit search \citep{2006MNRAS.373..799C} of the resulting 30,260 
photometric measurements found a strong 3.04 day periodicity. The discovery 
light curve is displayed in Figure~\ref{fig:phot}a, folded on this period.

\begin{figure}
\centering                     
\includegraphics[height=8.4cm,angle=270]{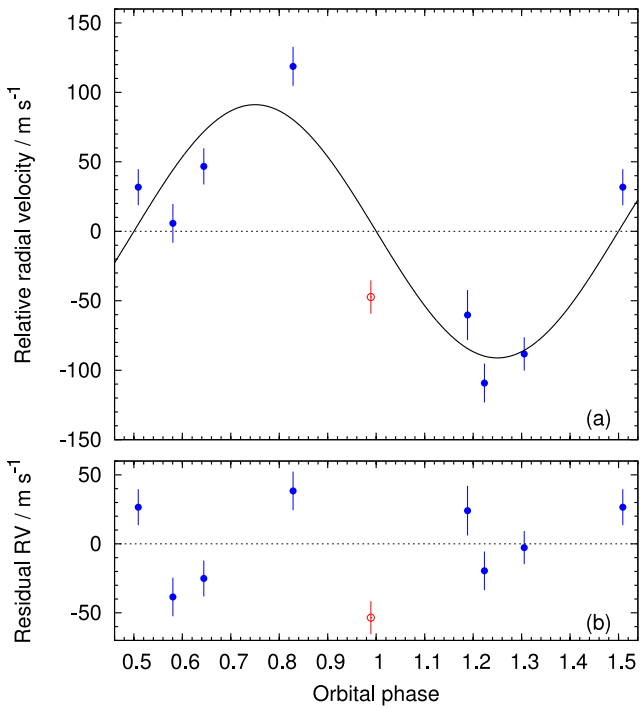}
\caption{{\bfseries(a)} SOPHIE RVs with the best-fitting circular 
Keplerian orbit superimposed.
The RV represented by an open, red circle was excluded from the analysis, as it 
was taken during transit. 
{\bfseries (b)} Residuals of the RVs about the fit.
\label{fig:rv}}
\end{figure}

Using the SOPHIE spectrograph mounted on the 1.93 m OHP telescope 
\citep{2008SPIE.7014E..17P,2009A&A...505..853B}, we obtained eight spectra of 
WASP-40 during 2010 April and May. 
The high-efficiency mode and slow readout were used, corresponding to a spectral 
resolving power of 40\,000 and lowest readout noise. We acquired a spectrum in 
both entrance fibers of the spectrograph to allow monitoring of and 
correcting for the sky background. However, over the sequence, this background 
was low enough that we did not have to apply such a correction. Signal-to-noise 
ratio values range from 22 to 35, for exposure times of 30 to 43 minutes. 
Radial-velocity (RV) measurements were computed by weighted cross-correlation 
\citep{1996A&AS..119..373B,2005Msngr.120...22P} with a numerical G2 spectral 
template. To account for systematic effects associated with the high-efficiency 
mode \citep[e.g.,][]{2009A&A...505..853B}, we added an uncertainty of 10 \ms\ in 
quadrature to the formal errors. 
RV variations were detected with the same period found from the transits 
and with a semiamplitude of 91 \ms, consistent with a planetary mass 
companion. The RV measurements are listed in Table~\ref{tab:rv} and are plotted 
in Figure~\ref{fig:rv}. 

\begin{table} 
\caption{SOPHIE Radial-Velocity Measurements} 
\label{tab:rv} 
\begin{tabular*}{0.5\textwidth}{@{\extracolsep{\fill}}cccc} 
\hline 
${\rm BJD}-2,400,000^{\rm a}$ & RV & $\sigma$$_{\rm RV}$ & BS\\ 
 & (km s$^{-1}$) & (km s$^{-1}$) & (km s$^{-1}$)\\ 
\hline
55,301.4903 & $-$15.796 & 0.012 & $-$0.035\\
55,303.4817 & $-$15.702 & 0.013 & $-$0.043\\
55,305.4918 & $-$15.837 & 0.012 & $-$0.044\\
55,323.4792 & $-$15.858 & 0.014 & $-$0.059\\
55,324.5650 & $-$15.743 & 0.014 & $-$0.025\\
55,334.4374 & $-$15.630 & 0.014 & $-$0.037\\
55,335.5319 & $-$15.809 & 0.018 & $-$0.054\\
55,336.5056 & $-$15.717 & 0.013 & $-$0.014\\
\hline 
\multicolumn{4}{l}{$^{\rm a}$ BJD = Barycentric Julian Date (UTC).}
\end{tabular*} 
\end{table} 

\begin{figure}
\centering                     
\includegraphics[width=8.4cm]{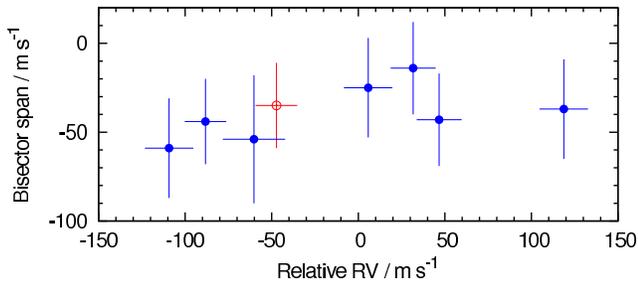}
\caption{  
Bisector span variation with respect to radial velocity, with approximately the 
same scale used for each axis. The systemic velocity 
was subtracted from the radial-velocity values. We adopted uncertainties on the 
bisector spans that were twice the size of those on the radial velocities.
\label{fig:bisector}}
\end{figure}

To test the hypothesis that the RV variations are due to spectral line 
distortions caused by a blended eclipsing binary or starspots, we performed a 
line-bisector analysis \citep{2001A&A...379..279Q} of the SOPHIE 
cross-correlation functions. 
The lack of correlation between bisector span and RV 
(Fig.~\ref{fig:bisector}) supports our conclusion that the periodic dimming of 
WASP-40 and its RV variations are due to a planet.

To refine the system parameters we obtained high-precision transit photometry. 
On 2010 June 26 we obtained a full transit of WASP-40 with the RISE 
(rapid imager to search for exoplanets) high-speed 
CCD camera mounted on the 2.0 m Liverpool Telescope 
\citep{rise2008, gibson2008}. 
RISE has a wideband filter of $\sim 500$--$700$ nm, which corresponds approximately 
to $V+R$. We obtained 285 exposures in the $2 \times 2$ binning mode with an
exposure time of 35 s and effectively no dead time. 
To minimize the impact of flat-fielding errors and to increase the duty cycle 
we defocused the telescope by 1 mm. 
On 2010 June 30 we observed a full transit of WASP-40 through a Gunn $r$ filter 
with the C2 camera on the 1.2 m Euler Swiss telescope, which was defocused by 
0.1 mm. The seeing ranged from 0.5 to 1.1\arcsec, and air mass ranged from 1.2 to 
1.5 over the course of the 105 exposures. 
We performed differential photometry relative to several stable reference stars, 
using the ULTRACAM pipeline \citep{Ultracam} for the RISE data and IRAF for 
the Euler data. 
The resulting light curves are displayed in Figure~\ref{fig:phot}c. 

\section{Stellar parameters from spectra}
The individual SOPHIE spectra of WASP-40 were co-added to produce a
single spectrum with an average S/N of around 60:1. The analysis was performed 
using the methods given in \citet{2009A&A...496..259G}. 
The H$\alpha$ line was used to determine the
effective temperature (\teff), and the Na {\sc i} D and Mg {\sc i} $b$ lines
were used as surface gravity (\logg) diagnostics. The parameters obtained from
the analysis are listed in Table~\ref{tab:stellar}. The elemental abundances
were determined from equivalent-width measurements of several clean and
unblended lines. A value for microturbulence (\mictrb) was determined from
Fe~{\sc i} using the method of \cite{1984A&A...134..189M}. The quoted error 
estimates include those given by the uncertainties in \teff, \logg\ and \mictrb, 
as well as the scatter due to measurement and atomic data uncertainties.

\begin{table}
\centering
\caption{Stellar Parameters from Spectra}
\begin{tabular}{lc}
\hline
Parameter & Value \\
\hline
\teff      & 5200 $\pm$ 150 K \\
\logg      & 4.5 $\pm$ 0.2 (cgs)\\
\mictrb    & 0.9 $\pm$ 0.2 \kms \\
\vsini     & 2.5 $\pm$ 0.9 \kms \\
{[Fe/H]}   & 0.14 $\pm$ 0.11\\
log A(Li)  &   $<0.5$  \\
\mstar     &  0.91 $\pm$ 0.08 \msol \\
\rstar     &  0.88 $\pm$ 0.22 \rsol \\ 
\hline
R.A. (J2000)	& 14$\rm^{h}$51$\rm^{m}$04.19$\rm^{s}$\\
Decl. (J2000)	& $+05^\circ$56$^{'}$50.5\arcsec\\
$m_V$	& $12.2 \pm 0.1$\\
$m_J$$\rm ^a$	& $10.63 \pm 0.03$\\
$m_H$$\rm ^a$	& $10.25 \pm 0.02$\\
$m_K$$\rm ^a$	& $10.11 \pm 0.02$\\
USNO-B1.0	& 0959-0237786\\
2MASS$\rm ^a$ 	& 14510418+0556505\\
\hline
\end{tabular}
\label{tab:stellar}
\\
$\rm ^a$ \citet{2006AJ....131.1163S}.
\end{table}

We estimated the sky-projected stellar rotation velocity (\vsini) by fitting the 
profiles of several unblended Fe~{\sc i} lines. For this, we used an 
instrumental FWHM of 
0.15 $\pm$ 0.01 \AA, determined from the telluric lines around 6300\AA\ and 
assumed a value for macroturbulence (\mactrb) of 1.0 $\pm$ 0.3 \kms\ 
\citep{2010MNRAS.405.1907B}. 

As a check of our analysis we also estimated the metallicity and \vsini\ 
directly from the cross-correlation function of the averaged SOPHIE spectra, 
using the methods described in \citet{2010A&A...523A..88B}. These give  
\feh\ = $0.06 \pm 0.09$ and \vsini\ = $3.3 \pm 1.0$ \kms, which are in agreement 
with our preceding values.

We input our values of \teff, \logg\ and \feh\ into the calibrations of 
\citet{2010A&ARv..18...67T} to obtain estimates of the stellar mass and radius 
(Table~\ref{tab:stellar}).

\section{System parameters from RV and transit data}
\label{sec:mcmc}
We determined the system parameters from a simultaneous fit to the data 
described in \S~\ref{sec:obs}.
The transit light curve was modeled using the formulation of 
\citet{2002ApJ...580L.171M} with the assumption that  \rplanet$\ll$\rstar. 
Limb-darkening was accounted for using a four-coefficient nonlinear limb-darkening 
model, using fixed coefficients (Table~\ref{tab:ld}) appropriate to the 
passbands and interpolated in effective temperature, surface gravity and 
metallicity from the tabulations of 
\citet{2000A&A...363.1081C}. 

\begin{table}
\caption{Limb-Darkening Coefficients} 
\label{tab:ld} 
\begin{tabular}{lcccc}
\hline
Light curve (band)	& $a_1$		& $a_2$		& $a_3$		& $a_4$	\\
\hline
WASP/Euler ($RC$)	& 0.714 	& $-$0.651 	& 1.335		& $-$0.602 \\
RISE ($VJ$)		& 0.657		& $-$0.626 	& 1.479		& $-$0.658 \\
\hline
\end{tabular}
\end{table}

The simultaneous fit was performed using the current version of the 
Markov-chain Monte Carlo (MCMC) code described by \citet{2007MNRAS.380.1230C} 
and \citet{2008MNRAS.385.1576P}. 
The transit light curve is parameterized by the epoch of midtransit 
$T_{\rm 0}$, the orbital period $P$, the planet-to-star area ratio 
(\rplanet/\rstar)$^2$, the approximate duration of the transit from initial to 
final contact $T_{\rm 14}$, and the impact parameter $b = a \cos i/R_{\rm *}$ 
(the distance, in fractional stellar radii, of the transit chord from the 
star's center). 
The radial-velocity orbit is parameterized by the stellar reflex velocity 
semiamplitude $K_{\rm *}$, the systemic velocity $\gamma$, 
and \secos\ and \sesin\ 
\citep{2011ApJ...726L..19A}, where $e$ is orbital 
eccentricity and $\omega$ is the argument of periastron. 

The linear scale of the system depends on the orbital separation $a$ which, 
through Kepler's third law, depends on the stellar mass \mstar. 
At each step in the Markov chain, the latest values of stellar density 
\densstar, effective temperature \teff\ and metallicity \feh\ are input 
to the empirical mass calibration of \citet{2010A&A...516A..33E} to obtain 
\mstar. 
The shapes of the transit light curve \citep{2003ApJ...585.1038S} and the 
radial-velocity curve constrain \densstar, which combines with \mstar\ to give 
\rstar. 
\teff\ and \feh\ are proposal parameters constrained by Gaussian priors with 
mean values and variances derived directly from the stellar spectra 
(Table~\ref{tab:stellar}). 

As the planet-star area ratio is constrained by the measured transit depth, 
\rplanet\ follows from \rstar. The planet mass \mplanet\ is calculated from 
the measured value of $K_1$ and 
\mstar; the planetary density \densplanet\ and surface gravity $\log g_{\rm P}$ 
then follow. 
We also calculate the blackbody equilibrium temperature $T_{\rm P, A=0}$, where 
$A$ is albedo, assuming efficient redistribution of heat from the planet's 
presumed permanent day side to its night side. 

At each step in the MCMC procedure, model transit light curves and 
radial-velocity curves are computed from the proposal parameter values, which are 
perturbed from the previous values by a small, random amount. The $\chi^2$ 
statistic is used to judge the goodness of fit of these models to the data and a 
step is accepted if $\chi^2$ is lower than for the previous step. A step 
with higher $\chi^2$ is accepted with a probability $\exp(-\Delta \chi^2$/2). 
To give proper weighting to each transit and RV data set, the uncertainties are 
scaled at the start of the MCMC so as to obtain a reduced $\chi^2$ of unity. 

From an initial MCMC fit for an eccentric orbit, we found 
$e = 0.13^{+0.18}_{-0.10}$. 
The improvement in the fit resulting from the addition of \secos\ and \sesin\ as 
fitting parameters is too small to justify adoption of an eccentric orbit. 
The $F$-test approach of \citet{1971AJ.....76..544L} indicates 
that there is an 84 \% probability that the improvement in the fit could 
have arisen by chance if the underlying orbit were circular. In the absence of 
conclusive evidence to the contrary, we adopted the circular orbit model.

One spectrum of the eight SOPHIE spectra was taken during the start of transit. 
As we did not fit for the Rossiter-McLaughlin effect 
\citep[e.g.,][]{2000A&A...359L..13Q} we excluded the resulting RV 
measurement (BJD = 2,455,301.4903) from our analysis. 

The median values and 1 $\sigma$ uncertainties of the system parameters derived 
from the MCMC model fit are presented in Table~\ref{tab:mcmc}. 
The corresponding transit and orbit models are superimposed on the transit 
photometry and radial velocities in Figures~\ref{fig:phot} and \ref{fig:rv}.

\begin{table} 
\caption{System Parameters from RV and Transit Data} 
\label{tab:mcmc} 
\begin{tabular*}{0.5\textwidth}{@{\extracolsep{\fill}}lc} 
\hline 
Parameter & Value \\ 
\hline 
$P$ & $3.0395589 \pm 0.0000090$ days\\
$T_{\rm c}$$^{\rm a}$ & HJD $2,455,362.31489 \pm 0.00023$\\
$T_{\rm 14}$ & $0.0696 \pm 0.0011$ days\\
$T_{\rm 12} \approx T_{\rm 34}$$^{\rm b}$ & 0.0282$^{+ \rm undefined}_{- 0.0037}$ days\\
$\Delta F=R_{\rm P}^{2}$/R$_{*}^{2}$ & 0.0152$^{+ 0.00041}_{- 0.00057}$\\
$b$ & 0.866$^{+ 0.016}_{- 0.012}$\\
$i$  \medskip & 85.01$^{+ 0.20}_{- 0.26}$ $^\circ$\\
$K_{\rm 1}$ & $91 \pm 13$ \ms\\
$\gamma$ & $-15748.7 \pm 1.0$ \ms\\
$e$ \medskip & 0 (adopted)\\

\mstar & $0.921 \pm 0.034$ \msol\\
\rstar & $0.864 \pm 0.031$ \rsol\\
\logg & $4.529 \pm 0.027$ (cgs)\\
\densstar & $1.43 \pm 0.13$ \denssol\\
\teff\ & $5246 \pm 153$ K\\
\feh\ \medskip & $0.13 \pm 0.11$\\

\mplanet & $0.617 \pm 0.088$ \mjup\\
\rplanet & 1.038$^{+ 0.068}_{- 0.050}$ \rjup\\
$\log g_{\rm P}$ & $3.112 \pm 0.080$ (cgs)\\
\densplanet\ & $0.54 \pm 0.12$ \densjup\\
$a$ & $0.03995 \pm 0.00050$ AU\\
$T_{\rm P, A=0}$ & $1177 \pm 42$ K\\
\hline 
\end{tabular*} 
$^{\rm a}$ HJD = Heliocentric Julian Date (UTC).\\
$^{\rm b}$ 1 $\sigma$ upper limit undefined as system is near-grazing.
\end{table}

\begin{figure}
\centering                     
\includegraphics[height=8.4cm,angle=270]{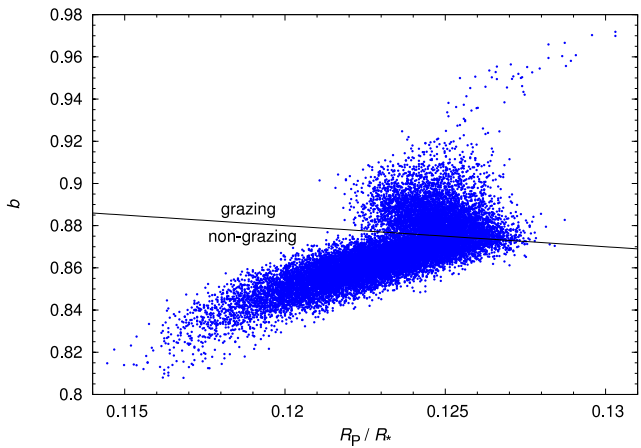}
\caption{  
The MCMC posterior distributions of $b$ and \rplanet/\rstar. The 
solid line indicates the position of the stellar limb, 
i.e., $b$ + \rplanet/\rstar = 1. 
A total of 26.4\% of the points lie above the line and are grazing 
solutions. 
\label{fig:graz}}
\end{figure}

For a transit to be grazing, the {\it grazing criterion} 
\citep{2011A&A...526A.130S} must be satisfied: 
\begin{equation}
X = b+{R_{\rm P}}/{R_{\star}} > 1.
\end{equation}
For WASP-40b $X = 0.9895^{+0.017}_{-0.014}$. 
Figure~\ref{fig:graz} shows the MCMC posterior distribution of 
\rplanet/\rstar\ and $b$. 
A total of 26.4\% of these data points satisfy the {\it grazing criterion}. 
Using the {\it odds ratio test} \citep[e.g.,][]{2010ApJ...725.2017K}, we find 
a 40.5\% probability that the system is grazing. 

\section{System age}
Assuming aligned stellar-spin and planetary-orbit axes, the measured \vsini\ of 
WASP-40 and its derived stellar radius (Table~\ref{tab:stellar}) indicate a 
rotational period of $P_{\rm rot} = 17.8 \pm 7.8$ days. Combining this with the 
$B - V$ color of a
K0 star from \citet{2008oasp.book.....G}, we used the relationship of 
\citet{2007ApJ...669.1167B} to estimate a gyrochronological age of 
$1.2^{+1.3}_{-0.8}$ Gyr. 
Considering that the stellar-spin axis may not be in the sky plane, these are 
upper limits on the stellar rotation period and the gyrochronological age. 
We found no evidence for rotational modulation in the WASP light curves. 

\begin{figure}
\centering                     
\includegraphics[width=8.4cm]{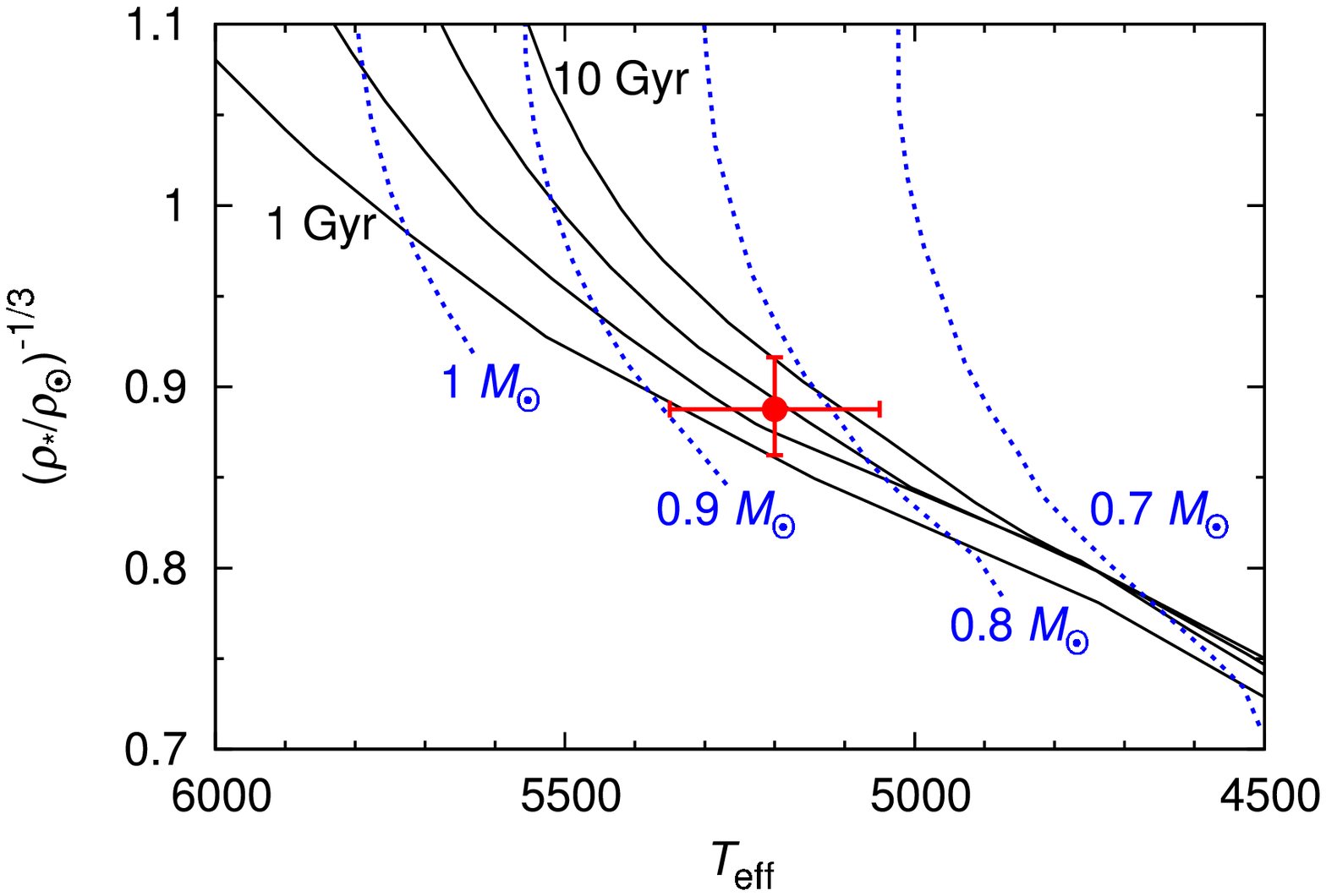}
\caption{  
Modified H-R diagram. The evolutionary mass tracks ($Z = 0.026 \approx$ 
\feh\ = 0.14; $Y = 0.30$) are from \citet{2008A&A...484..815B}. The 
isochrones ($Z = 0.026\approx$ \feh\ = 0.14) 
for the ages 1, 4, 7, and 10 Gyr are from \citet{2008A&A...482..883M}.
\label{fig:evol}}
\end{figure}

We interpolated the stellar evolution tracks of \citet{2008A&A...482..883M} and 
\citet{2008A&A...484..815B} 
using \densstar\ from the MCMC analysis and using \teff\ and \feh\ from the 
spectral analysis (Fig.~\ref{fig:evol}). 
This suggests an age of $6 \pm 5$ Gyr and a mass of $0.83 \pm 0.07$ \msol\ for 
WASP-40.

\section{Discussion}
As we find more exoplanets that transit their host stars
we begin to see patterns
in their distribution.  It is thus important to add new systems
to probe such patterns and to understand the role of selection
effects in transit surveys.  For example, given the apparent
pileup of transiting planets with periods near 3--4 days, it is
important to investigate the shape of the cutoff toward 
shorter periods \citep[e.g.,][]{2011ApJ...727L..44S}.  Further, there
are increasing suggestions of correlations between planetary
radii, the irradiation of the planet and the metallicity
of the host star \citep{2011MNRAS.410.1631E,2011inprep...prad}.  
For brighter systems,
the measurement of the Rossiter--McLaughlin effect allows
us to build up statistics of the alignment of planetary
orbits \citep[e.g.,][]{2010A&A...524A..25T}, and to thus test suggestions
of a correlation between alignment and the spectral type
of the host star \citep{2010ApJ...718L.145W}.

Since all such comparisons depend on the reliability of
measured parameters of the transiting systems, it is
worth noting that the parameters of HAT-P-27b/WASP-40b
reported here are in good accord with those from the
independent study by \citet{2011arXiv1101.3511B}. Most parameters
agree to within 1 $\sigma$ errors, while the impact parameter,
which is strongly dependent on the modeling of the
transit light curve and the limb-darkening, agrees to within 
2 $\sigma$.  This adds confidence to the methods used
by the two projects.

\begin{figure}
\centering                     
\includegraphics[width=8.4cm]{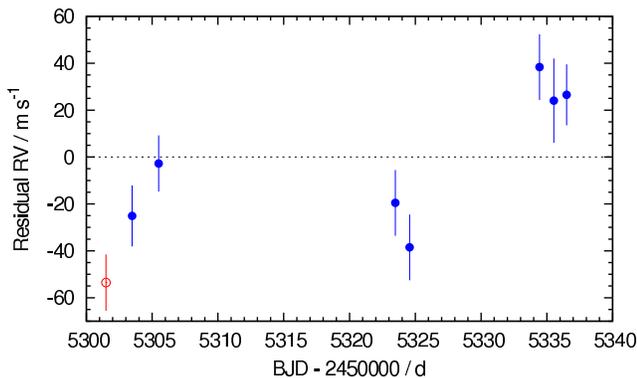}
\caption{Residual radial velocities about the best-fitting circular Keplerian 
orbit.
\label{fig:rv-time}}
\end{figure}

One difference in our analyses is that our radial velocities 
show more scatter about the model than would be expected 
from their formal uncertainties (Fig.~\ref{fig:rv-time}). 
This could be an indication of stellar activity.  
We calculated an activity index suggestive of an active star, 
$\log R'_{\rm HK}$ = $-4.63$, directly from the SOPHIE spectra using the 
methods described in \citet{2010A&A...523A..88B}. 
The RV dispersion arising from the stellar activity of a K-type star is 
estimated to be $\sim$10 \ms\ \citep{2000A&A...361..265S}.

We note that the 
radial-velocity data in \citet{2011arXiv1101.3511B}, while resulting in a very 
similar model, do not show a similar scatter, which could indicate that the 
level of stellar activity fluctuates. 
Alternatively, if the scatter was indicative of an additional planet, the signal 
may not be present in the radial velocities of \citet{2011arXiv1101.3511B} due 
to limited sampling. 
Both the SOPHIE and Keck data sets are sparse, with eight and nine radial 
velocities, respectively, so more observations are needed to reach a conclusion.

\acknowledgments
The research leading to these results has received funding from the European 
Community's Seventh Framework Programme (FP7/2007-2013) under grant agreement 
number RG226604 (OPTICON). 
SuperWASP-N is hosted by the Issac Newton Group on La Palma and WASP-South is 
hosted by the South African Astronomical Observatory. 
We are grateful for their ongoing support and assistance. 
Funding for WASP comes from consortium universities and from the UK's Science 
and Technology Facilities Council.
We thank Tom Marsh for the use of the ULTRACAM pipeline.

{\it Facilities:} SuperWASP, OHP:1.93m, Liverpool:2m, Euler1.2m


\end{document}